\documentclass[a4paper,12pt]{article}

\usepackage{color,graphicx}
\usepackage{amsmath,amsfonts,amssymb,amsthm,mathrsfs}
\usepackage{epsf,epsfig,subfigure,psfrag,graphicx,afterpage}
\usepackage{chicago}

\begin{document}

\pagestyle{plain} \vspace{1in}

\title{Bayesian GARMA Models for Count Data}

\author{
  Marinho G. Andrade, Ricardo S. Ehlers, Breno S. Andrade
}

\date{}

\maketitle

\begin{abstract}
  
Generalized autoregressive moving average (GARMA) models are a class
of models that was developed for extending the univariate Gaussian
ARMA time series model to a flexible observation-driven model for
non-Gaussian time series data. This work presents Bayesian approach
for GARMA models with Poisson, binomial and negative binomial
distributions. A simulation study was carried out to investigate the
performance of Bayesian estimation and Bayesian model selection
criteria. Also three real datasets were analysed using the Bayesian approach
on GARMA models. 
\vskip .5cm


\end{abstract}

\section{Introduction}

Observed counts as time series have been attracting considerable
attention both in terms of data analysis and developement of
methodological approaches. This type of data can appear in contexts as
diverse as Epidemiology (see for example \citeNP{Zeger} and
\shortciteNP{Davis2}) and Finance (\shortciteNP{Liesenfeld} and
\citeNP{Rydberg}). In this paper, the motivating datasets that will be
analyzed are the number of automobile production in Brazil, the number
of hospitalizations caused by Dengue Fever and the number of 
deaths in Brazil.

Parameter and observation driven models provide a flexible framework
for modelling time series of counts. So far, a wide variety of models
for count time
series have been discussed in the literature usually embedded in the
framework of integer valued
ARMA type models (see for example \citeNP{biswas}). An overview of
these kind of models can be found in \citeN{Davis2} while
\citeN{Zeger} and \citeN{Chan} explicitly discuss and develop
estimation techniques for Poisson generalized linear models with an
autoregressive latent process in the mean. 

\shortciteN{Davis1} proposed a flexible framework for modelling a wide
range of dependence structures using models for Poisson
counts. \shortciteN{jung} compares various models for time series of counts
which can account for discreteness, over dispersion and serial
correlation. \citeN{Zhu} proposed a negative binomial INGARCH model
applied to the Polio data discussed in \citeN{Zeger}. 

This article extends the work of \shortciteN{Benjamin}, giving rise to the
Bayesian approach on the generalized autoregressive moving average
(GARMA) model. This approach presents some gain in terms of estimation,
that could be more adequate using different loss functions. The use
of Bayesian selection criteria is also an import contribution from this
article. Last but not least the application of discrete
models on important Brazilian real data providing a new perspective on
this field. 

The remainder of this paper is organized as follows. Section \ref{sec:garma}
defines the GARMA model 
with discrete distributions. The Bayesian approach and Bayesian
prediction are presented in Section \ref{sec:bayes}. Section \ref{simulations} describes the
simulation study where the performance of the Bayesian approach for
estimation and selection was investigated.
Real data applications are illustrated on Section 5. Finally, Section
\ref{conclusions} gives some concluding remarks. 

\section{Generalized Autoregressive Moving Average Model}\label{sec:garma}

The GARMA model, introduced by \shortciteN{Benjamin}, assumes that the
conditional distribution of each observation $y_t$, for $t=1,\dots,n$
given the previous information set $\textit{F}_{t-1} =
(x_{1},\ldots,x_{t-1},y_{1},\ldots,y_{t-1},\mu_{1},\ldots,\mu_{t-1})$
belongs to the exponential family. The conditional density is given by,
\begin{equation}\label{defgarma}
f(y_t|\textit{F}_{t-1})=
\exp\left(\frac{y_t\alpha_t - b(\alpha_t)}{\varphi} + d(y_t,\varphi)\right),
\end{equation}
where $\alpha_t$ e $\varphi$ are conical and scale parameter
respectively, with $b(\cdot)$ e $d(\cdot)$ being specific functions
that define the particular exponential family. The conditional mean
and conditional variance of $y_t$ given $\textit{F}_{t-1}$ is
represented by the terms $\mu_t  = E(y_t|\textit{F}_{t-1}) =
b'(\alpha_t)$ and $Var(y_t|\textit{F}_{t-1}) =
\varphi{b''}(\alpha_t)$, with $t=1,\ldots,n$.

Just as in Generalized Linear Models (GLM, \citeNP{McCullagh}), $\mu_t$,
is related to the linear predictor, $\eta_t$, by a twice-differentiable
one-to-one monotonic link function $g(\cdot)$. The linear predictor for
the GARMA model is given by,
\begin{equation}\label{predii33}
g(\mu_t) = \eta_t = x'_t\beta + \sum_{j=1}^p\phi_j\{g(y_{t-j}) -
x'_{t-j}\beta\} + \sum_{j=1}^q\theta_j\{g(y_{t-j}) - \eta_{t-j}\}.
\end{equation}

The GARMA($p$,$q$) model is defined by equations (\ref{defgarma}) and
(\ref{predii33}). For certain functions $g$, it may be necessary to
replace $y_t$ with $y_t^{*}$ in (\ref{predii33}) to avoid the
non-existence of $g(y_t)$ for certain values of $y_t$. The form
$y_t^{*}$ depends on the particular function $g(.)$ and is defined for
specific cases later.

The definition of GARMA model allows to consider the adjust of
exogenous variables $x'_t$ however in this work the term
$x'_t\beta$ will be considered as a constant $\beta_0$. For count data
time series we will consider the following distributions. 

\subsection{Poisson GARMA model}

Suppose that $y_t|\textit{F}_{t-1}$ follows a Poisson distribution
with mean $\mu_t$. Then,
\begin{equation}\label{eqi}
f(y_t|\textit{F}_{t-1}) = \exp\left\{y_t\log(\mu_t) - \mu_t  - \log(y_{t}!) \right\}.
\end{equation}
and $Y_t|\textit{F}_{t-1}$ has distribution in the exponential family
with $\varphi = 1$, $\alpha_t=\log(\mu_t)$, 
$b(\alpha_t)=\exp(\alpha_t)$, $c(y_t,\varphi)=-\log(y_{t}!)$ and 
$\nu(\mu_t)=\mu_t$. 
The canonical link function for this model is the logarithmic
function, so that the linear predictor is given by,
\begin{equation}\label{mediagarma1}
\log(\mu_t) = \beta_0 + \sum_{j=1}^{p}\phi_j\{\log{y_{t-j}^*}\} +
\sum_{j=1}^{q}\theta_j\{\log(y^*_{t-j}) - \log(\mu_{t-j})\}, 
\end{equation}
Where $y_{t-j}^* = \max(y_{t-j},c), 0<c<1$. The Poisson GARMA model is
defined by equations (\ref{eqi}) and (\ref{mediagarma1}). 

\subsection{Binomial GARMA model}

Suppose that $y_t|\textit{F}_{t-1}$ follows a binomial distribution
with mean $\mu_t$. Then,
\begin{footnotesize}
\begin{equation*}\label{eqiqq}
f(y_t|\textit{F}_{t-1}) = 
\exp\left\{  {y_t}\log\left( \frac{\mu_t}{m-\mu_t} \right) +
m\log\left( \frac{m-\mu_t}{m} \right) + \log\left(
{\frac{\Gamma(m+1)}{\Gamma(y_t + 1)\Gamma(m - y_t + 1)}} \right)
\right\}. 
\end{equation*} 
\end{footnotesize}

The canonical link function for this model is the logarithmic
function. The linear predictor is given by, 
\begin{equation}\label{mediagarma12}
\log\left( \frac{\mu_t}{m-\mu_t} \right) = \beta_0 +
\sum_{j=1}^{p}\phi_j\{\log{y_{t-j}^*}\} +
\sum_{j=1}^{q}\theta_j\{\log(y^*_{t-j}) - \log(\mu_{t-j})\}, 
\end{equation}
with $y_{t-j}^* = \max(y_{t-j},c), 0<c<1$, and $m$ is known.

\subsection{Negative Binomial}

Let $y_t$ a time series such that $y_t|\textit{F}_{t-1}\sim NB(k,\mu_t)$. Then,
\begin{equation*}\label{bngarma}
f(y_t|\textit{F}_{t-1}) =
\exp\left(k\log\left\{\frac{k}{\mu_t+k}\right\} +
y_t\log\left\{\frac{\mu_t}{\mu_t+k}\right\} + \log\left\{
\frac{\Gamma(k+y_t)}{\Gamma(y_{t}+1)\Gamma(k)} \right\}\right), 
\end{equation*}
which belongs to the exponential family with $k$ known.
The link function for this model is the logarithmic function
\begin{equation*}\label{mediagarma}
\log\left(\frac{k}{\mu_t+k}\right) = \beta_0 +
\sum_{j=1}^{p}\phi_j\{\log{y_{t-j}^*}\} +
\sum_{j=1}^{q}\theta_j\{\log(y^*_{t-j}) - \log(\mu_{t-j})\}, 
\end{equation*}
with $y_{t-j}^* = \max(y_{t-j},c), 0<c<1$. 

\section{Bayesian Approach on GARMA Models}\label{sec:bayes}

\subsection{Defining the Prior Densities}

Using the logarithmic in link
function to guarantee positive values for any values of the vectors
$\boldsymbol{\beta} = (\beta_1,\ldots,\beta_m)$, 
$\Phi=(\phi_1,\ldots,\phi_p)$ and 
$\Theta=(\theta_1,\ldots,\theta_q)$.$\beta$, $\phi_i$. Thus, a multivariate
Gaussian prior will be proposed for each parameter. 
\begin{eqnarray*}
\boldsymbol{\beta} &\sim & N(\boldsymbol{\mu_0},\sigma_0^{2}\boldsymbol{I_0}),\\
\Phi &\sim & N(\boldsymbol{\mu_1},\sigma_1^{2}\boldsymbol{I_1})  \\
\Theta &\sim & N(\boldsymbol{\mu_2},\sigma_2^{2}\boldsymbol{I_2})\\
\end{eqnarray*}
where $\boldsymbol{\mu_0},\boldsymbol{\mu_1},\boldsymbol{\mu_1}$ are
vectors with length $m$, $p$ and $q$ respectively,
$\sigma_0^{2}$, $\sigma_1^{2}$ and $\sigma_1^{2}$ represent the prior
variance and $\boldsymbol{I_0}$, $\boldsymbol{I_1}$ and
$\boldsymbol{I_2}$ are $m\times m$, $p\times p$
and $q\times q$ identity matrices
respectively. The construction of the multivariate Gaussian depends on
hyper parameters, when there is no prior knowledge on these parameters
it can be considered a vary large variance making the prior densities
flats. The partial likelihood function for GARMA models can be
constructed as follows 

\begin{eqnarray*} \label{q1}
L(\boldsymbol{\beta},\Phi,\Theta|Y) &\propto& \prod_{t=r+1}^{n}f(y_t|F_{t-1})  \nonumber\\
                           &\propto&
\prod_{t=r+1}^{n}\exp\left(\frac{y_{t}\alpha_{t} -
  b(\alpha_{t})}{\varphi} + d(y_{t},\varphi)\right), 
\end{eqnarray*}
where $\alpha_t = g(\mu_t)$, which represent the link function given by

\begin{equation*}\label{q2}
g(\mu_t) = x'_t\boldsymbol{\beta} + \sum_{j=1}^p\phi_j\{g(y_{t-j}^*) -
x'_{t-j}\} + \sum_{j=1}^q\theta_j\{g(y_{t-j}^*) - g(\mu_{t-j})\}, 
\end{equation*}
for all $t = r+1, \ldots, n$.

The posterior density is obtained combining the likelihood
function with the prior densities. Let the vector $\boldsymbol{Y}=
(y_t,y_{t-1},\dots,y_1, x_t, x_{t-1},\dots,x_1,\dots)$
represent the necessary information to construct the likelihood
function. The posterior density is then given by,
\begin{equation}\label{jjj1}
\pi(\boldsymbol{\beta},{\Phi},{\Theta}|\textit{Y}) \propto
L(\boldsymbol{\beta},{\Phi},{\Theta}|\textit{Y})\pi_{0}(\boldsymbol{\beta},{\Phi},{\Theta}). 
\end{equation}

\noindent However, the joint posterior density of parameters in the
GARMA models can not 
be obtained in closed form. Therefore, Markov chain Monte Carlo (MCMC)
sampling strategies will be employed for obtaining samples from this
joint posterior distribution. In particular, we use a
Metropolis-Hastings algorithm to yield the required realisations. We
adopt a sampling scheme where the parameters are
updated as o single block and at each iteration we generate new values
from a multivariate normal distribution centred around the maximum
likelihood estimates
with a variance-covariance proposal matrix given by the inverse
Hessian evaluated at the posterior mode.

\subsection{Bayesian prediction on GARMA models}

An important aspect of our Bayesian approach to GARMA
models is the hability to forecasting future values of the time
series, $y_{t+h}$, $h\geq 1$ given all the information available until
time $t$.
To evaluate this forecasting it is necessary to find the predictive
density function $p(y_{t+h}|Y)$.

Denoting the information set
$\widehat{\textit{F}}_{t+h} = (\widehat{x}_{t+h},\dots
,x_{t},x_{t-1},\dots,$
$\widehat{y}_{t+h-1},\dots,y_{t},$ $y_{t-1},\dots$
$\widehat{\mu}_{t+h-1}$,$\dots,$ $\mu_{t},\mu_{t-1},\dots)$, where
$\widehat{y}_{t+h-i}=y_{t+h-i}$, if $h\leq i$, else
$\widehat{y}_{t+h-i}=E\{y_{t+h-i}|\widehat{\textit{F}}_{t+h-i} \}$,
$i=1,2,\ldots h+1$. The general idea is that
$\widehat{\textit{F}}_{t+h}$ contains all the data observed until the
time $t$, for the future time $t+h$, $h\geq 1$, the set
$\widehat{\textit{F}}_{t+h}$ is completed with forecasts of necessary
information to estimate $y_{t+h}$. Starting with, 

\begin{equation}\label{model}
f(y_{t+h}|\boldsymbol{\beta},{\Phi},{\Theta},\widehat{\textit{F}}_{t+h})
= \exp\left(\frac{y_{t+h}\alpha_{t+h} - b(\alpha_{t+h})}{\varphi} +
d(y_{t+h},\varphi)\right), 
\end{equation}

The conditional mean and variance of $y_{t+h}$ given
$\widehat{\textit{F}}_{t+h}$ is represented by the terms
$\widehat{\mu}_{t+h} = $ $E(y_{t+h}|\widehat{\textit{F}}_{t+h})$
$=b'(\alpha_{t+h})$ and $Var(y_{t+h}|\textit{F}_{t+h})=$
$\varphi{b''}(\alpha_{t+h})$. 
The $\mu_{t+h}$, is related to the predictor, $\eta_{t+h}$, by a
twice-differentiable one-to-one monotonic link function
$g(\cdot)$. The linear predictor for the GARMA model is given by, 

\begin{equation}\label{predii33}
g(\mu_{t+h}) = \eta_{t+h} = \widehat{x'}_{t+h}\beta +
\sum_{j=1}^p\phi_j\{g(\widehat{y}_{t+h-j}) -
\widehat{x'}_{t+h-j}\beta\} +
\sum_{j=1}^q\theta_j\{g(\widehat{y}_{t+h-j}) -
\widehat{\eta}_{t+h-j}\}. 
\end{equation}

With the equation \eqref{model} and posterior density \eqref{jjj1},
the predictive density for $y_{t+h}$ can be written as, 

\begin{equation*}\label{jjhhh3}
p(y_{t+h}|\widehat{F}_{t+h}) =
\int_{\{\boldsymbol{\beta},{\Phi},{\Theta}\} \in
  \Omega}f(y_{t+h}|\boldsymbol{\beta},{\Phi},{\Theta},\widehat{F}_{t+h})\pi(\boldsymbol{\beta},{\Phi},{\Theta}|
Y) d\boldsymbol{\beta} d\Phi d\Theta. 
\end{equation*}

The aim is to determine the predictive density using the MCMC algorithm, thus

\begin{equation}\label{jjhhh5}
\widehat{p}(y_{t+h}|\widehat{\textit{F}}_{t+h}) =
\frac{1}{Q}\sum_{j=1}^{Q}f(y_{t+h}|\boldsymbol{\beta}^{(j)},{\Phi^{(j)}},{\Theta^{(j)}},\widehat{\textit{F}}_{t+h}). 
\end{equation}

Given the predictive density, the next step is to evaluate the prediction,
$E(y_{t+h}|\widehat{\textit{F}}_{t+h}) = \hat{y}_{t+h}$.

\begin{equation}\label{jjhhh7}
E(y_{t+h}|\widehat{\textit{F}}_{t+h}) = 
\int_{y_{t+h} \in R}y_{t+h}p(y_{t+h}|\widehat{\textit{F}}_{t+h})dy_{t+h}.
\end{equation}

\noindent Substituting the equation \eqref{jjhhh5} the equation (\ref{jjhhh7}) can be rewritten by,

\begin{eqnarray}\label{jjhhh8}
&&
  E(y_{t+h}|\widehat{\textit{F}}_{t+h}) = \nonumber\\
&&
\int_{y_{t+h} \in R}y_{t+h}
\left[ \int_{\{\boldsymbol{\beta},{\Phi},{\Theta}\} \in
    \Omega}f(y_{t+h}|\boldsymbol{\beta},{\Phi},{\Theta},\widehat{\textit{F}}_{t+h})\pi(\boldsymbol{\beta},{\Phi},{\Theta}|
  Y)d\boldsymbol{\beta}d\Phi d\Theta\right] dy_{t+h}.\nonumber\\ 
\end{eqnarray}

\noindent Using properties of integer, we can rewrite \eqref{jjhhh8} as,

\begin{eqnarray*}\label{jjhhh9}
&&
E(y_{t+h}|\widehat{\textit{F}}_{t+h}) = \\
&&
\int_{\{\boldsymbol{\beta},{\Phi},{\Theta}\} \in \Omega} \left[
  \int_{y_{t+h} \in R}y_{t+h}
  f(y_{t+h}|\boldsymbol{\beta},{\Phi},{\Theta},\widehat{\textit{F}}_{t+h})dy_{t+h}\right]
\pi(\boldsymbol{\beta},{\Phi},{\Theta}| Y)d\boldsymbol{\beta} d\Phi
d\Theta,
\end{eqnarray*}
which can in turn be rewritten as
\begin{equation*}\label{jjhhh10}
E(y_{t+h}|\widehat{\textit{F}}_{t+h}) =
\int_{\{\boldsymbol{\beta},{\Phi},{\Theta}\} \in \Omega} \left[
  E(y_{t+h}|\boldsymbol{\beta}
  ,{\Phi},{\Theta},\widehat{\textit{F}}_{t+h}) \right]
\pi(\boldsymbol{\beta} ,{\Phi},{\Theta}| Y)d\boldsymbol{\beta} d\Phi
d\Theta. 
\end{equation*}

\noindent Now, denoting 
$\mu_{t+h}(\boldsymbol{\beta},\Phi,\Theta,\widehat{\textit{F}}_{t+h})=
E(y_{t+h}|\boldsymbol{\beta},\Phi,\Theta,\widehat{\textit{F}}_{t+h})$ and
using the MCMC output vector
$(\boldsymbol{\beta}^{(j)},\Phi^{(j)},\Theta^{(j)})$,
$j=1,2,\ldots,Q$, it follows that\linebreak $E(y_{t+h}|\widehat{\textit{F}}_{t+h})$ can be
approximated by,
\begin{equation*}\label{jjhhh11}
\widehat{y}_{t+h} =
\frac{1}{Q}\sum_{k=1}^Q\mu_{t+h}(\boldsymbol{\beta}^{(k)},\Phi^{(k)},\Theta^{(k)},\widehat{\textit{F}}_{t+h}), 
\end{equation*}
where
\begin{eqnarray}\label{predii33}
&& g(\mu_{t+h}^{(k)})= \nonumber\\
&& \widehat{x'}_{t+h}\boldsymbol{\beta}^{(k)} +
\sum_{j=1}^p\phi_j^{(k)}\{g(\widehat{y}_{t+h-j}) -
\widehat{x'}_{t+h-j}\boldsymbol{\beta}^{(k)}\} +
\sum_{j=1}^q\theta_j^{(k)}\{g(\widehat{y}_{t+h-j}) -
\widehat{\eta}_{t+h-j}^{(k)}\}.\nonumber\\ 
\end{eqnarray}

Credible intervals for $\widehat{y}_{t+h}$ can be calculated using
the $100\alpha\%$, and $100(1-\alpha)\%$ quantiles of the MCMC
sample $\mu^{(k)}_{t+h}$, with $k=1,\ldots,Q$. An approach to estimate
the credible interval of $\widehat{y}_{t+h}$ is the Highest
Posterior Density (HPD), see \citeN{hpd}.
A $100(1-\alpha)\%$ HPD region for $\widehat{y}_{t+h}$ are a subset $C
\in R$ defined by $C = \{y_{t+h} :
p(y_{t+h}|\widehat{\textit{F}}_{t+h})\geq \kappa\}$, where $\kappa$ is
the largest number such that 
\begin{equation}
\int_{y_{t+h} \geq \kappa}p(y_{t+h}|\widehat{\textit{F}}_{t+h})dy_{t+h}=1-\alpha.
\end{equation}
We can use the $\widehat{p}(y_{t+h}|\widehat{\textit{F}}_{t+h})$ MCMC
estimates, given by the equation \eqref{jjhhh5}, to estimate the
$100(1-\alpha)\%$ HPD region. 
We used the following algorithm to calculate the credible intervals
for the predictions.
\begin{itemize}
\item[1.] Let a sequence of forecast values $\widehat{y}_{t+h}$ for $h=1,\ldots,H$.
\item[2.] Take $h=1$, $k=0$, $y_{t+h}^{(0)}=0$, $S_{t+h}^{(0)}=0$ and also initiate $LB=0$, $UB$=0.
\item[3.] Using the initial values evaluate the equation:
\begin{equation*}
f(y_{t+h}^{(k)}|\beta^{(j)},\Phi^{(j)},\Theta^{(j)},\widehat{\textit{F}}_{t+h})
= \exp\left(\frac{y_{t+h}^{(k)}\alpha_{t+h}^{(j)} -
  b(\alpha_{t+h}^{(j)})}{\varphi} + d(y_{t+h}^{(k)},\varphi)\right), 
\end{equation*}
and also,

\begin{equation*}
\widehat{p}(y_{t+h}^{(k)}|\widehat{\textit{F}}_{t+h}) =
\frac{1}{Q}\sum_{j=1}^{Q}f(y_{t+h}^{(k)}|\beta^{(j)},{\Phi^{(j)}},{\Theta^{(j)}},\widehat{\textit{F}}_{t+h}). 
\end{equation*}

\item[4.] Using $\widehat{p}(y_{t+h}^{(k)}|\widehat{\textit{F}}_{t+h})$ compute $S_{t+h}^{(k+1)}$ with

\begin{equation*}
S_{t+h}^{(k+1)}=S_{t+h}^{(k)}+\widehat{p}(y_{t+h}^{(k)}|\widehat{\textit{F}}_{t+h})
\end{equation*}

\item[5.] If $LB=0$ and $S_{t+h}^{(k+1)}\geq\delta$, $\rightarrow$ $y_{t+h,\delta}=y_{t+h}^{(k)}$ and $LB=1$.
\item[6.] If $UB=0$ and $S_{t+h}^{(k+1)}\leq(1-\delta)$, $\rightarrow$ $y_{t+h,(1-\delta)}=y_{t+h}^{(k)}$ and $UB=1$.
\item[7.] If $LB=0$ or $UB=0$, take $k=k+1$ and
  $y_{t+h}^{(k)}=y_{t+h}^{(k-1)}+1$, repeat steps 3 and 4 until $LB=1$
  and $UB=1$. 

\end{itemize}

The percentiles $100\delta\%$ and $100(1-\delta)\%$ are represented by
$y_{t+h,\delta}$ and $y_{t+h,(1-\delta)}$ respectively, and given by,

\begin{equation*}
y_{t+h,\delta}=\max\left\{y_{t+h}^{(r)} \big| \sum_{k=1}^{r}
\widehat{p}(y_{t+h}^{(k)}\big|\widehat{\textit{F}}_{t+h})\leq \delta
\right\}. 
\end{equation*}

\begin{equation*}
y_{t+h,(1-\delta)}=\min\left\{y_{t+h}^{(r)}\big| \sum_{k=1}^{r}
\widehat{p}(y_{t+h}^{(k)}\big|\widehat{\textit{F}}_{t+h})\geq
(1-\delta) \right\}. 
\end{equation*}
and the $100(1-\delta)\%$ credible interval for the predictions is
denoted by $CI_{(1-\delta)}=\left[y_{t+h,\delta};y_{t+h,(1-\delta)}\right]$.

\section{Simulation Study}\label{simulations}

In this section we conduct a simulation study for negative binomial\linebreak
GARMA($p,q$) models with different orders $p$ and $q$.
The actual parameter values used to simulate the artificial series are shown in
Table \ref{tab1} and the parameter $k$ of the negative binomial was
fixed at $k=15$. These values were chosen taking into account that a
GARMA model can be nonstationary since they are in the exponencial
family and the variance function depends on the mean. So, we opted to
chose parameter values that would generate moderate values for the time series.
The experiment was replicated $m=1000$ times for each model. For each
dataset we used the prior distributions as described in Section
\ref{sec:bayes} with mean zero and variance 200. We then drew samples
from the posterior distribution 
discarding the first 1000 draws as burn-in and keeping every 3rd
sampled value resulting in a final sample of 5000 values.
All the computations were implemented
using the open-source statistical software language and environment {\tt R}
\cite{r10}.

\begin{table}[h]
\caption{Parameters values to simulate from Negative Binomial GARMA($p$,$q$).}
\label{tab1}
\begin{center}
\renewcommand{\arraystretch}{1.2}
\begin{scriptsize}
\begin{tabular}{|c|c|c|c|c|c|c}
\hline
Order & $\beta_0$ & $\phi_1$ &  $\phi_2$  & $\theta_1$ & $\theta_2$ \\
\hline
(1,1) & 0.80      & 0.50     &  -        & 0.30       & -          \\
(1,2) & 1.00      & 0.30     &  -        & 0.40       & 0.25       \\
(2,1) & 0.55      & 0.30     &  0.40     & 0.20       & -          \\
(2,2) & 0.65      & 0.30     &  0.40     & 0.25       & 0.35       \\
\hline
\end{tabular}
\end{scriptsize}
\end{center}
\end{table}

The performance of the Bayesian estimation was evaluated using three
metrics: the corrected bias (CB), the corrected error (CE) and the
mean acceptance rates in the MCMC algorithm called Acceptance
Probabilities (AP). These metrics are defined as,
\begin{eqnarray*}
CB &=& \frac{1}{m} \sum_{i=1}^m 
\left|\frac{\theta-\hat{\theta}^{(i)}}{\theta}\right|,\label{bias}\\
CE^2 &=& 
\frac{1}{Var}\frac{1}{m}\sum_{i=1}^m(\hat{\theta}^{(i)}-\theta)^2\label{mse}\\
AP &=& \frac{1}{m} \sum_{i=1}^m \hat{r}^{(i)}\label{ap},
\end{eqnarray*}
where $\hat{\theta}^{(i)}$  and $\hat{r}^{(i)}$ are the estimate of
parameter $\theta$ and the computed acceptance rate respectively for
the $i$-th replication, $i=1,\ldots,m$. In this paper we take the
posterior means of $\theta$ as point estimates. Also, the variance
term ($Var$) that appears in the definition of CE is the sample
variance of $\hat{\theta}^{(1)},\dots,\hat{\theta}^{(m)}$.

The estimation results appear in Table \ref{tab2} where the posterior
mean and variance (in brackets) as well as the aforementioned metrics
are shown for each model and parameter. These results indicate good
properties with relatively small values of the corrected bias (CB),
values of the corrected error (CE) around 1 and acceptance
probabilities between 0.20 and 0.70.

We also include Table \ref{ga3} with the 
proportions of correct model choice using three popular Bayesian model selection
criteria. Specifically, we adopt the expected Bayesian information criterion 
(EBIC, \shortciteNP{carl00}), the Deviance information criterion (DIC,
\shortciteNP{Spiegelhalter}) and the conditional predictive ordinate 
(CPO, \shortciteNP{gelfand}) to
select the order of the GARMA models. Each column in this table
contains the model order and the associated proportions of correct
model choice according to EBIC, DIC and CPO criteria. Higher
proportions of correct model choices are observed as the sample sizes
increase for all models and criteria. Also, EBIC and CPO tend to
perform better for GARMA(1,1) and GARMA(1,2) models but none performed
particularly well with GARMA(2,2) models.

Finally, this simulation study was carried out also for the Poisson and
binomial distributions with results similar to the ones shown. These
results are not included to save space.

\setlength{\tabcolsep}{1mm}

\begin{table}[h]
\caption{Monte Carlo experiments. Corrected bias, corrected errors and
  mean acceptance rates for the Bayesian estimation of Negative
  Binomial GARMA($p$,$q$) model.}
\label{tab2}
\begin{center}
\renewcommand{\arraystretch}{1.2}
\begin{scriptsize}
\begin{tabular}{c|cccc|cccc}
\hline
Parameter & Mean(Var)(1,1)& CB(1,1)& CE(1,1)& AP(1,1)& Mean(Var)(1,2)& CB(1,2)& CE(1,2)& AP(1,2)\\
\hline
$\beta_0$ & 0.8571(0.0065)& 0.0984 & 1.2247 & 0.3746 & 1.0823(0.0196)& 0.1276 & 1.1592 & 0.3182 \\
$\phi_1$  & 0.4695(0.0026)& 0.0947 & 1.1637 & 0.3511 & 0.2554(0.0097)& 0.2820 & 1.0965 & 0.2702 \\
$\phi_2$  & -             & -      & -      & -      & -             & -      & -      & -      \\
$\theta_1$& 0.2927(0.0033)& 0.1531 & 1.0071 & 0.6480 & 0.4099(0.0091)& 0.1900 & 1.0048 & 0.4327 \\
$\theta_2$& -             & -      & -      & -      & 0.2478(0.0037)& 0.1929 & 1.0001 & 0.5882 \\
\hline
\end{tabular}

\begin{tabular}{c|cccc|cccc}
\hline
Parameter & Mean(Var)(2,1) & CB(2,1)& CE(2,1)& AP(2,1)& Mean(Var)(2,2) & CB(2,2)& CE(2,2)& AP(2,2)\\
\hline
$\beta_0$ & 0.6198(0.0097) & 0.1740 & 1.2240 & 0.2786 & 0.7344(0.0079) & 0.1497 & 1.3171 & 0.3397\\
$\phi_1$  & 0.2798(0.0152) & 0.3295 & 1.0127 & 0.1422 & 0.2887(0.0054) & 0.1959 & 1.0111 & 0.2282\\
$\phi_2 $ & 0.3794(0.0066) & 0.1661 & 1.0307 & 0.2091 & 0.3414(0.0049) & 0.1485 & 1.0787 & 0.2348\\
$\theta_1$& 0.2012(0.0182) & 0.5334 & 0.9995 & 0.3214 & 0.2430(0.0052) & 0.2307 & 1.0040 & 0.5237\\
$\theta_2$& -              & -      & -      & -      & 0.3464(0.0027) & 0.1193 & 1.0017 & 0.6614\\ 
\hline
\end{tabular}
\end{scriptsize}
\end{center}
\end{table}

\begin{table}[ht!]
\caption{Proportions of correct model chosen via Bayesian criteria with
 Negative Binomial GARMA($p$,$q$) models.} 
\label{ga3}
\begin{scriptsize}
\begin{center}
\renewcommand{\arraystretch}{1.2}
\begin{tabular}{ccccc}
  \hline
\multicolumn{5}{c}{\textbf{EBIC}}\\
\hline
Size  & GARMA(1,1) &  GARMA(1,2) & GARMA(2,1) &  GARMA(2,2)\\
\hline  
200   & 0.9379     &  0.3042     & 0.5626     &  0.4450    \\
500   & 0.9799     &  0.6156     & 0.8048     &  0.5825    \\
1000  & 0.9852     &  0.9039     & 0.8471     &  0.6772    \\
\hline
\end{tabular}

\begin{tabular}{ccccc}
  \hline
\multicolumn{5}{c}{\textbf{DIC}}\\
\hline
Size    & GARMA(1,1) & GARMA(1,2)  &  GARMA(2,1) & GARMA(2,2)\\
\hline  
 200    & 0.6316     & 0.4804      &  0.5445     & 0.4437    \\
 500    & 0.6876     & 0.6476      &  0.6221     & 0.4925    \\
1000    & 0.7155     & 0.7364      &  0.6469     & 0.7154    \\ 
\hline
\end{tabular}  

\begin{tabular}{ccccc}
  \hline
\multicolumn{5}{c}{\textbf{CPO}}\\
\hline
Size    & GARMA(1,1) & GARMA(1,2) &  GARMA(2,1) &  GARMA(2,2) \\
\hline 
 200    & 0.8078     & 0.3493     &  0.5575     &  0.4112     \\
 500    & 0.8188     & 0.5925     &  0.5993     &  0.4625     \\
1000    & 0.8325     & 0.7266     &  0.6152     &  0.7317     \\ 
\hline
\end{tabular}
\end{center}
\end{scriptsize}
\end{table}

\section{Bayesian Real Data Analysis}

In this section, we apply the methodology described so far to three
real time series of count data. For each series we estimated
GARMA($p,q$) models with varying orders and computed the Bayesian
selection criteria EBIC, DIC and CPO for model comparison. In all
cases we used the diagnostic proposed by \citeN{geweke} to assess
convergence of the chains. This is based on a test for equality
of the means of the first and last part of the chain (by default
the first 10$\%$ and the last 50$\%$). If the samples are drawn from
the stationary distribution, the two means are equal and
the statistic has an asymptotically standard normal
distribution. The calculed values of Geweke statistics were all
between -2 and 2 which is an indication of convergence of
the Markov chains.

\subsection{Automobile data set}

The first real data set analysed is the number of automobile
production in Brazil between January 1993 and December 2013. The data
is available from {\tt http://www.anfavea.com.br/tabelas.html}.
The original observations were divided by 1000 to reduce the magnitude
of the data.

\begin{figure}[h]\centering
\includegraphics[angle=270,width=0.8\linewidth]{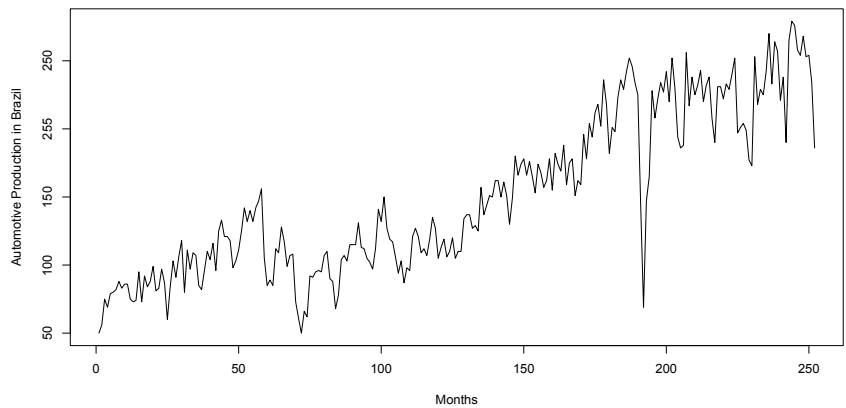}
\caption{Graph of number of automobile production in Brazil.}
\label{auto1}
\end{figure}

The automotive industry is extremely important as it can influence
other industries activities. For example, 50\% of the world rubber
production, 25\% of the world glass production and 15\% of the world iron
production are destined to the automotive industry. 
The behaviour of the data along time depicted in Figure \ref{auto1}
seems to indicate that an extra term should be included to take into
account a (possibly nonlinear) trend. The term $\beta_{{\exp}}=\log(t)$ was
then included in the model equation
to account for this long-term increase.

\setlength{\tabcolsep}{0.5mm}

\begin{table}[h]
\caption{Bayesian selection criteria for the number of automobile production in Brazil.}
\label{tab4}
\begin{center}
\renewcommand{\arraystretch}{1.2}
\begin{scriptsize}
\begin{tabular}{ccccccc}
\hline
Poisson  & GARMA(1,0)& GARMA(2,0) & GARMA(1,1) &  GARMA(1,2) &  GARMA(2,1) &  GARMA(2,2)\\
\hline  
EBIC     & 3046.61   & 3074.24    & 3045.70    &  3074.45    &  3071.21    &  3067.97   \\
DIC      & 3032.06   & 3064.97    & 3030.38    &  3064.55    &  3046.02    &  3065.89   \\
CPO      &-1519.88   &-1536.12    &-1519.65    & -1535.15    & -1536.76    & -1540.29   \\
\hline
Binomial & GARMA(1,0)& GARMA(2,0) & GARMA(1,1) &  GARMA(1,2) &  GARMA(2,1) &  GARMA(2,2)\\
\hline  
EBIC     & 3559.79   & 3814.33    &  3559.11   &  3813.91    &  3759.32    &  3738.38   \\ 
DIC      & 3545.12   & 3736.57    &  3544.19   &  3794.01    &  3738.90    &  3713.19   \\
CPO      &-1782.36   &-1930.30    & -1780.67   & -1949.77    & -1929.13    & -1909.67   \\
\hline
Negative Binomial& GARMA(1,0) &  GARMA(2,0)  &\textbf{GARMA(1,1)}  & GARMA(1,2) & GARMA(2,1) & GARMA(2,2)\\
\hline
EBIC     &  2547.67  &  2792.38   &\textbf{2546.76}   &  2799.21    & 2787.56    & 2785.10  \\
DIC      &  2537.71  &  2777.16   &\textbf{2531.85}   &  2779.28    & 2767.32    & 2760.13  \\
CPO      & -1269.48  & -1427.72   &\textbf{-1267.34}  & -1430.66    &-1426.47    &-1423.09  \\ 
\hline
\end{tabular}
\end{scriptsize}
\end{center}
\end{table}

The results regarding selection criteria are summarized in Table
\ref{tab4}. We note that the three criteria indicate that the most appropriate
model was the GARMA(1,1) Negative Binomial. 
Also, Table \ref{tab5}
presents the estimation results for the selected GARMA(1,1) Negative
Binomial model with the extra parameter fixed at $k=150$.

\begin{table}[h]
\caption{Estimation results. GARMA(1,1) Negative
Binomial model for number of automobile production in Brazil.}
\label{tab5}
\begin{center}
\begin{footnotesize}
\renewcommand{\arraystretch}{1.2}
\begin{tabular}{ccccc}
\hline
Parameter     & Mean   & Variance & HPD Credible Interval &  AP \\\hline  
$\beta_0$     & 0.3834 & 0.0006   & (0.3543; 0.4159) & 0.3710 \\
$\beta_{{\exp}}$& 0.0850 & 0.0002  & (0.0814; 0.0884) & 0.3163 \\ 
$\phi_1$      & 0.8447 & 0.0005   & (0.8379; 0.8521) & 0.3038 \\
$\theta_1$    & 0.1149 & 0.0005   & (0.1064; 0.1244) & 0.6323 \\   
\hline
\end{tabular}
\end{footnotesize}
\end{center}
\end{table}

\begin{figure}[h]\centering
\includegraphics[angle=270,width=0.9\linewidth]{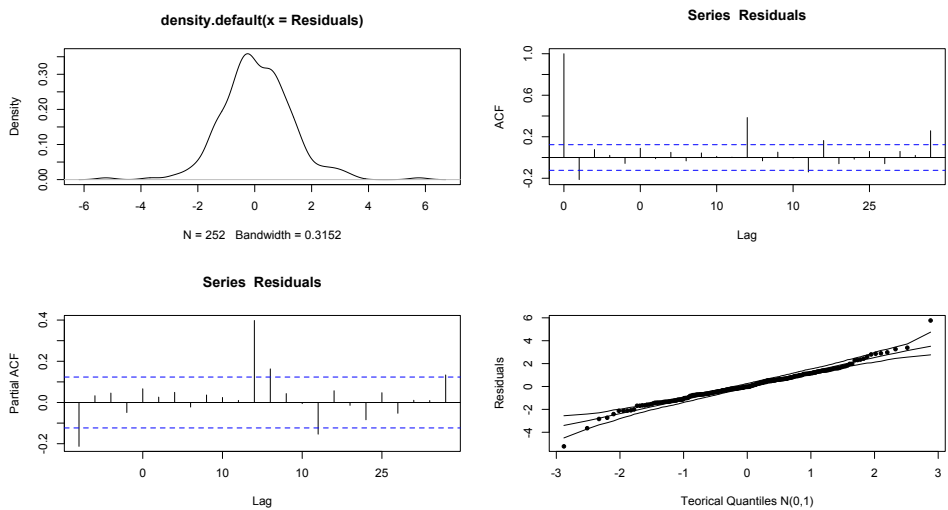}
\caption{Residual Analysis for the number of automobile production in
  Brazil under a GARMA(1,1) negative binomial model.}
\label{fig:res1}
\end{figure}

We also performed a residual analysis based on the so called quantile
residuals which are the common choice for generalized linear models. 
In fact, quantile residuals are the only
useful residuals for binomial, negative binomial or Poisson data when
the response takes on only a small number of distinct values (\citeNP{dunn}).
These are given by 
$r_t=\Phi^{-1}(\textbf{F}_{y_t}(y_t|F_{t-1}))$ where
$\textbf{F}_{y_t}$ represent the cumulative distribution function of
the associated discrete distribution. In practice, when dealing with
discrete distributions we need to introduce some randomization to
produce continuous normal residuals.
The residual analysis summarized in Figure \ref{fig:res1} which
indicates that the residuals are non-correlated
and Gaussian distributed with mean 0.0767 and standard deviation 1.2295. 
Kolmogorov-Smirnov and Lilliefors normality tests returned $p$-values
of 0.4502 and 0.0743 respectively which provides evidence for
Gaussian assumption (\citeNP{kolmo1}). 

\begin{figure}[h]\centering
\includegraphics[angle=270,width=0.7\linewidth]{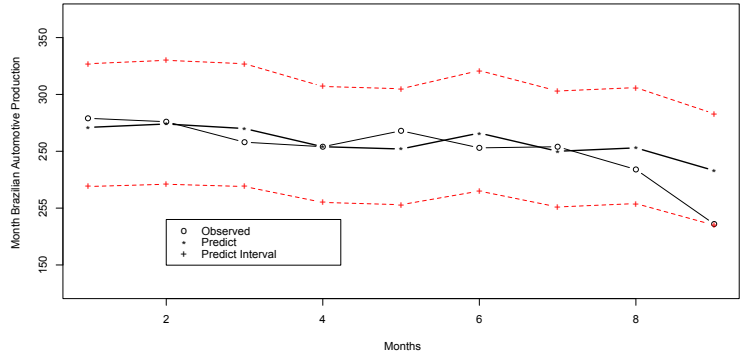}
\caption{Predictions for the number of automobile production in Brazil
  with a GARMA(1,1) Negative Binomial model.}
\label{fig:pred1}
\end{figure}

Finally, we performed a prediction exercise using the last 9
observations of the original series as follows. For each $k=1,\dots,9$ the
GARMA(1,1) negative binomial model was fitted to the series
$y_1,\dots,y_{n-k}$ and an out-of-sample one-step ahead prediction
$\hat{y}_{n-k+1}$ was produced. These predictions can then be compared
with the true values.
The results are illustrated in Figure \ref{fig:pred1} from which we
can see that the prediction errors are overall small.
A formal comparison was made by calculating the mean absolute percentage
error (MAPE, \citeNP{hyndman}) and we obtained the value $6.07\%$.

\subsection{Epidemiology data set}

This real data set comprises the number of hospitalizations caused by Dengue Fever
in Campina Grande city (Brazil) between January 1998 and October 2003.  
Dengue Fever is transmitted by several species of mosquito within the $genus
Aedes$, principally {\it A. aegypti}. The {\it Aedes} mosquito is easily
identifiable by the distinctive black and white stripes on its
body. It prefers to lay eggs on clean and stagnant water. Analysing the
autocorrelation function of this data, a seasonal behaviour is characterised. 
This is because the Summer months in this region present higher volume
of rain, thus leading to more clean and
stagnant water. Therefore we included two seasonal components in the model,
$\beta_{{S}_1}$ and $\beta_{{S}_2}$, using {\tt cosine} and
{\tt sine} functions
respectively, and also considering the period of 12 months. These
components are expected to improve model estimation.

\begin{figure}[h!]
\centering
\includegraphics[angle=270,width=0.8\linewidth]{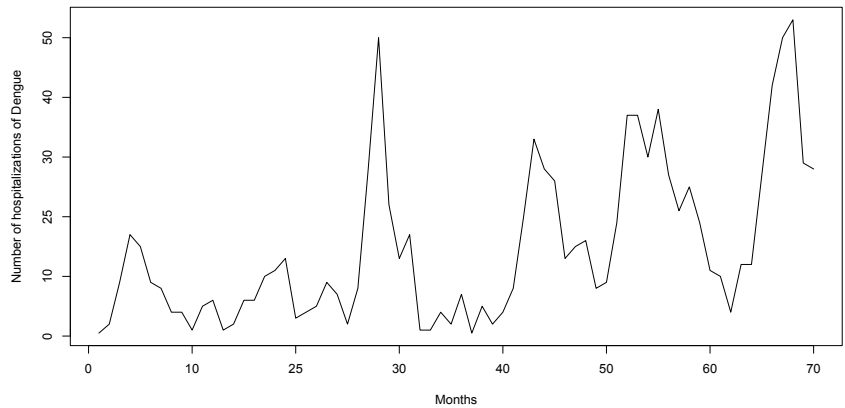}
\caption{Number of hospitalizations caused by Dengue Fever.}
\end{figure}

\setlength{\tabcolsep}{0.5mm}

\begin{table}[h]
\caption{Bayesian selection criteria for the number of hospitalizations
  caused by Dengue Fever.}
\label{tab6}
\begin{center}
\renewcommand{\arraystretch}{1.2}
\begin{scriptsize}
\begin{tabular}{ccccccc}
\hline
Poisson & GARMA(1,0)& GARMA(2,0)& GARMA(1,1)& GARMA(1,2)   & GARMA(2,1) & GARMA(2,2) \\
\hline  
EBIC    & 632.82    & 633.13    & 633.73    & 632.48       & 632.65     & 628.20     \\
DIC     & 580.66    & 581.04    & 581.86    & 581.25       & 580.32     & 578.31     \\
CPO     &-794.03    &-794.87    &-794.69    &-794.11       &-793.83     &-792.34     \\
\hline
Binomial& GARMA(1,0)& GARMA(2,0)& GARMA(1,1)& GARMA(1,2)   & GARMA(2,1) & GARMA(2,2) \\
\hline
EBIC    & 690.62    & 689.28    & 690.34    & 656.56       & 688.82     & 655.30  \\
DIC     & 679.14    & 679.92    & 679.42    & 642.12       & 674.83     & 637.19  \\
CPO     &-345.89    &-346.16    &-345.76    &-327.13       &-348.04     &-324.83  \\
\hline
Negative Binomial   & GARMA(1,0)& GARMA(2,0)& GARMA(1,1)   &\textbf{GARMA(1,2)} & GARMA(2,1) & GARMA(2,2)\\
\hline  
EBIC    & 507.89    & 508.97    & 509.36    &\textbf{504.12} & 509.09     & 505.89\\
DIC     & 519.66    & 520.93    & 520.22    &\textbf{518.30} & 523.11     & 523.19\\ 
CPO     &-256.35    &-255.88    &-256.10    &\textbf{-254.24}&-257.64     &-256.26\\ 
\hline
\end{tabular}
\end{scriptsize}
\end{center}
\end{table}

The results regarding the selection criteria are summarized in Table
\ref{tab6} from which we can conclude that the most appropriate
model was the GARMA(1,2) Negative Binomial. Note that the three
criteria gave the same indication.
Table \ref{tab7} shows the estimation results for the selected GARMA(1,2) Negative
Binomial model with the extra parameter fixed at $k=30$.

\begin{table}[h]
\caption{Estimation results. GARMA(1,2) negative binomial model for
  the number of hospitalizations caused by Dengue Fever.} 
\label{tab7}
\begin{center}
\begin{footnotesize}
\renewcommand{\arraystretch}{1.2}
\begin{tabular}{cccll}
\hline
Parameter   & Mean    &  Variance  &HPD Credible Interval & AP \\
\hline  
$\beta_0$   & 1.1916  &  0.0566    & ( 0.7443; 1.6068) & 0.1090\\
$\beta_{S_1}$&-0.2571  &  0.0035    & (-0.3753;-0.1407) & 0.6196\\
$\beta_{S_2}$& 0.1424  &  0.0040    & ( 0.0156; 0.2649) & 0.5858\\
$\phi_1$    & 0.5796  &  0.0078    & ( 0.4230; 0.7456) & 0.0968\\
$\theta_1$  & 0.1214  &  0.0112    & (-0.0853; 0.3273) & 0.3391\\
$\theta_2$  & 0.0987  &  0.0053    & (-0.0470; 0.2358) & 0.3978\\   
\hline
\end{tabular}
\end{footnotesize}
\end{center}
\end{table}

\begin{figure}[h!]\centering
\includegraphics[angle=270,width=0.7\linewidth]{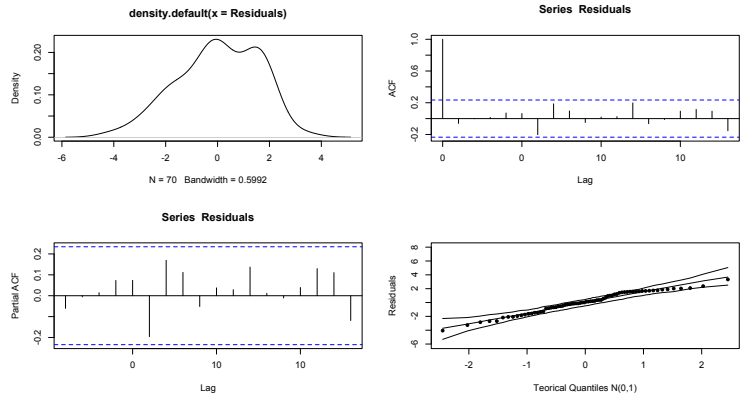}
\caption{Residual Analysis of Hospitalizations caused by Dengue.}
\label{fig:res2}
\end{figure}

Again we performed a residual analysis based on quantile
residuals. This is summarized in Figure \ref{fig:res2} which indicates
that the residuals are non-correlated 
and Gaussian distributed with mean 0.0258 and standard deviation 1.5571.
The Kolmogorov-Smirnov and Shapiro-Wilk normality tests returned
$p$-values of 0.4856 and 0.1176 respectively thus giving evidence for
the Gaussian assumption.

\begin{figure}[h!]
\centering
\includegraphics[angle=270,width=0.7\linewidth]{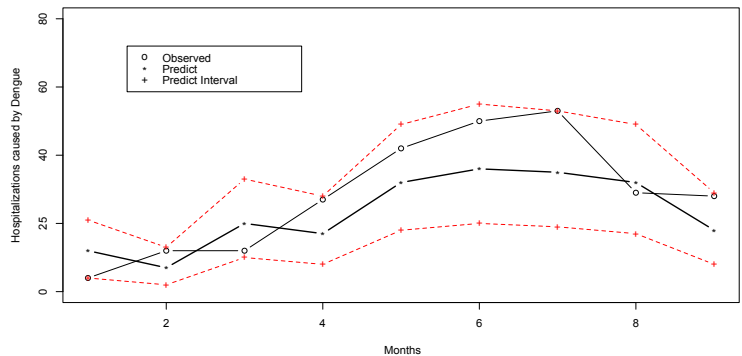}
\caption{Predictions with GARMA(1,2) Negative Binomial model with Hospitalizations caused by Dengue series.}
\label{fig:pred2}
\end{figure}

A similar prediction exercise was performed for this data. So, we
fitted a GARMA(1,2) negative binomial model to $y_1,\dots,y_{n-k}$ and
computed an out-of-sample one-step ahead prediction $\hat{y}_{n-k+1}$ for
$k=1,\dots,9$. 
Figure \ref{fig:pred2} shows the predictions,
prediction intervals and the real observations for comparison. It can
be seen that, although relatively close to the actual values,
predictions for May, June, July and August 2003
are consistently below the
observations. The MAPE criterion was calculated as $47.81\%$.

\subsection{Mortality data set}

\begin{figure}[h!]
\centering
\includegraphics[angle=270,width=0.8\linewidth]{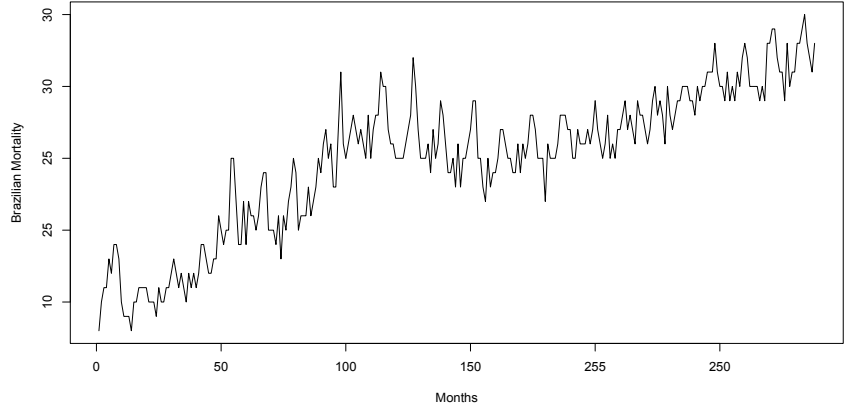}
\caption{Number of deaths in Brazil.}
\label{fig:deaths}
\end{figure}

Our last real data set is the number of deaths in Brazil between January
1984 and December 2007. This data is available from the Brazilian Health
Ministry at {\tt http://www2.datasus.gov.br/DATASUS}
and is depicted in Figure \ref{fig:deaths}. Likewise the
first example, the original series was divided by 1000 to reduce the
magnitude of the data.
As in the first example, we think there is a point for the inclusion
of an extra term here too since the series exhibits a long-term
(possibly nonlinear)
increase. So, a new component $\beta_{{\exp}}=\log(t)$ was added to
the model equation as this is expected to improve model estimation.

\begin{table}[h]
\caption{Bayesian selection criteria using the number of deaths in Brazil.}
\label{tab8}
\begin{center}
\begin{scriptsize}
\renewcommand{\arraystretch}{1.2}
\begin{tabular}{ccccccc}
\hline
Poisson &  GARMA(1,0)&  GARMA(2,0) & GARMA(1,1)  &  GARMA(1,2)  & GARMA(2,1)& GARMA(2,2) \\ 
\hline  
EBIC    &  1549.55   &  1560.41    & 1546.60     &  1566.37     & 1565.79   & 1566.80    \\ 
DIC     &  1531.53   &  1566.68    & 1531.10     &  1570.34     & 1571.11   & 1573.77    \\ 
CPO     & -766.42    & -773.35     & -765.49     & -784.20      & -784.99   &-785.48     \\ 
\hline
Binomial& \textbf{GARMA(1,0)}&  GARMA(2,0) & GARMA(1,1) &  GARMA(1,2) &  GARMA(2,1) &  GARMA(2,2) \\ 
EBIC    & \textbf{1351.42}   &  1412.95    & 1357.52    &  1391.79    &  1399.13    &  1404.81  \\ 
DIC     & \textbf{1341.42}   &  1391.56    & 1342.28    &  1371.54    &  1378.10    &  1379.88  \\ 
CPO     & \textbf{-670.73}   & -705.64     & -671.10    & -695.29     & -716.72     & -708.52   \\ 
\hline
Negative Binomial & GARMA(1,0) & GARMA(2,0)   & GARMA(1,1) & GARMA(1,2) & GARMA(2,1) & GARMA(2,2) \\
\hline  
EBIC         & 1705.33  & 1709.12  & 1700.61  &  1735.23   & 1734.39  & 1738.30  \\
DIC          & 1693.59  & 1696.61  & 1685.18  &  1714.76   & 1713.51  & 1712.47  \\
CPO          &-851.35   &-855.13   &-842.01   & -866.51    &-866.45   &-866.47   \\
\hline
\end{tabular}
\end{scriptsize}
\end{center}
\end{table}

Looking at the Bayesian selection criteria given in Table \ref{tab8} we
can conclude that the best model for this particular data is the
GARMA(1,0) Binomial model. There are only three parameters in this model and
the estimation results are shown in Table \ref{tab9}. Here the extra
parameter was fixed at $m=45$. 

\begin{table}[h]
\caption{Estimates of the number of deaths in Brazil series with GARMA(1,0) Binomial.}
\label{tab9}
\begin{center}
\begin{footnotesize}
\renewcommand{\arraystretch}{1.2}
\begin{tabular}{ccccc}
\hline
Parameter     & Mean   & Variance &HPD Credible Interval& AP \\
\hline
$\beta_0$     & 0.4154 & 0.0006   & (0.3739; 0.4724) & 0.2272\\
$\beta_{\exp}$ & 0.0713 & 0.0004   & (0.0651; 0.0774) & 0.3503\\
$\phi_1$      & 0.7637 & 0.0007   & (0.7462; 0.7788) & 0.1885\\ 
\hline
\end{tabular}
\end{footnotesize}
\end{center}
\end{table}

\begin{figure}[h!]\centering
\includegraphics[angle=270,width=0.7\linewidth]{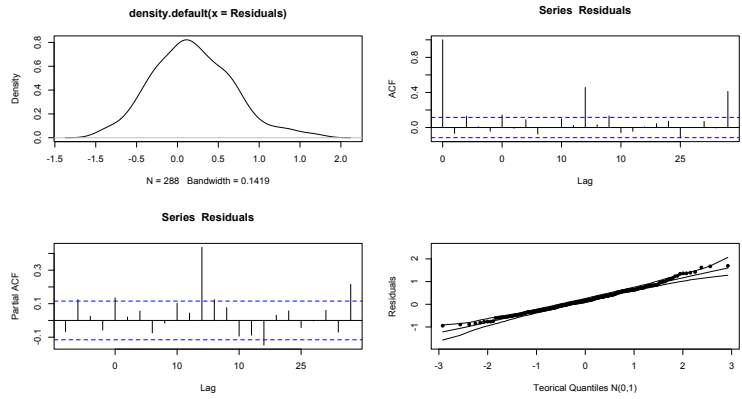}
\caption{Residual analysis of the number of deaths in Brazil.}
\label{fig:res3}
\end{figure}

The residual analysis summarized in Figure \ref{fig:res3} indicates
that the residuals are non-correlated 
and Gaussian distributed with mean 0.1850 and standard deviation 0.4894. The
Kolmogorov-Smirnov and Anderson-Darling normality tests returned
$p$-values of 0.6736 and 0.1304 respectively thus indicating evidence
for the Gaussian assumption. 

\begin{figure}[h!]\centering
\includegraphics[angle=270,width=0.7\linewidth]{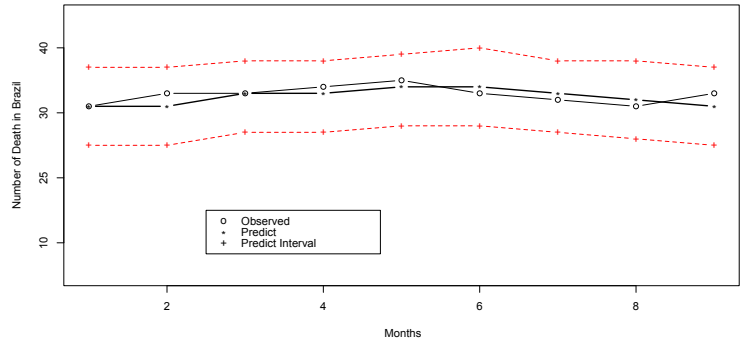}
\caption{Predictions with GARMA(1,0) Binomial model with Number of death in Brazil series.}
\label{fig:pred3}
\end{figure}

Likewise the previous examples we repeated the prediction exercise
here. This time we used the 10 last observations as the series is
longer. So, the GARMA(1,0) binomial model was fitted to the series
$y_1,\dots,y_{n-k}$ and a one-step ahead prediction $\hat{y}_{n-k+1}$
was produced for $k=1,\dots,10$.
The results are illustrated in Figure \ref{fig:pred3} from which we
can see that the prediction errors are again overall small.
Using these prediction errors the calculated value for the MAPE
criterion was $3.63\%$. 

\section{Discussion}\label{conclusions}

In this paper we discuss a Bayesian approach for estimation,
comparison and prediction of GARMA time series
models. We analysed three different discrete models:
Poisson, binomial and negative binomial. We implemented MCMC
algorithms to carry out the simulation study and the methodology was
also applied on three real discrete time series data.

Properties of the Bayesian estimation and the performance of Bayesian
selection criteria were assessed with our simulation study. The
analysis with real data also provided good estimates and predictions
via parsimonious models. All in all our results suggest that, as
indicated in the original GARMA paper, 
this class of models have potential uses for modelling overdispersed
time series count data.

\end{document}